\documentclass[aps,prl,twocolumn,superscriptaddress]{revtex4-1}

\usepackage{fancyhdr} 

\usepackage{hyperref} 

\usepackage{verbatim} 
\usepackage{amsmath} 
\usepackage{amsthm} 
\usepackage{amssymb}	
\usepackage{graphicx} 
\usepackage[dvips,letterpaper,margin=0.75in,bottom=0.5in]{geometry}

\let\baraccent=\= 
\renewcommand{\=}[1]{\stackrel{#1}{=}} 

\newcommand{\gae}{\lower 2pt \hbox{$\,
		\buildrel{\scriptstyle >}\over {\scriptstyle \sim}\,$}}
\newcommand{\lae}{\lower 2pt \hbox{$\,
		\buildrel{\scriptstyle <}\over {\scriptstyle \sim}\,$}}

\newcommand{\abs}[1]{\left| #1 \right|} 
\newcommand{\avg}[1]{\left< #1 \right>} 
 
\newcommand{\ket}[1]{\big| #1 \big\rangle} 
\newcommand{\braket}[2]{\big\langle #1 \vphantom{#2} \big|
	#2 \vphantom{#1} \big\rangle} 

\usepackage{color}
\usepackage{ulem}

\usepackage{pdfpages}

\makeatletter
\AtBeginDocument{\let\LS@rot\@undefined}
\makeatother

\begin{document}

\title{Floquet Dynamics of Boundary-Driven Systems at Criticality}

\author{William Berdanier}
\email[]{wberdanier@berkeley.edu}
\affiliation{Department of Physics, University of California, Berkeley, CA 94720, USA}

\author{Michael Kolodrubetz}
\affiliation{Department of Physics, University of California, Berkeley, CA 94720, USA}
\affiliation{Materials Sciences Division, Lawrence Berkeley National Laboratory, Berkeley, CA 94720, USA}

\author{Romain Vasseur}
\affiliation{Department of Physics, University of California, Berkeley, CA 94720, USA}
\affiliation{Materials Sciences Division, Lawrence Berkeley National Laboratory, Berkeley, CA 94720, USA}

\author{Joel E. Moore}
\affiliation{Department of Physics, University of California, Berkeley, CA 94720, USA}
\affiliation{Materials Sciences Division, Lawrence Berkeley National Laboratory, Berkeley, CA 94720, USA}

\date{\today}

\begin{abstract}
A quantum critical system described at low energy by a conformal field theory (CFT) and subjected to a time-periodic boundary drive displays multiple dynamical regimes depending on the drive frequency.  We compute the behavior of quantities including the entanglement entropy and Loschmidt echo, confirming analytic predictions from field theory by exact numerics on the transverse field Ising model, and demonstrate universality by adding non-integrable perturbations.  The dynamics naturally separate into three regimes: a slow-driving limit, which has an interpretation as multiple quantum quenches with amplitude corrections from CFT; a fast-driving limit, in which the system behaves as though subject to a single quantum quench; and a crossover regime displaying heating.  The universal Floquet dynamics in all regimes can be understood using a combination of boundary CFT and Kibble-Zurek scaling arguments. 
\end{abstract}
\maketitle


Recent years have witnessed substantial progress in understanding the dynamics of periodically driven (Floquet) systems. Such driving has traditionally been used for engineering non-trivial effective Hamiltonians~\cite{doi:10.1080/00018732.2015.1055918,PhysRevLett.116.205301,0953-4075-49-1-013001,PhysRevX.4.031027}, but recent research has shown that these dynamics can differ drastically from their static counterparts. Examples include the recently observed Floquet time crystals~\cite{khemani_phase_2016,pi-spin-glass,else_floquet_2016,else_pre-thermal_2016,zhang_observation_2016}, the emergence of topological quasiparticles protected by driving~\cite{jiang_majorana_2011,sreejith_parafermion_2016}, Floquet topological insulators~\cite{oka_photovoltaic_2009,kitagawa_topological_2010,lindner_floquet_2011,rechtsman_photonic_2013,cayssol_floquet_2013,titum_disorder-induced_2015}, and Floquet symmetry-protected topological phases~\cite{PhysRevB.92.125107,von_keyserlingk_phase_2016,PhysRevB.93.201103,potter_classification_2016}. More broadly, periodically driven systems touch on fundamental issues in statistical and condensed-matter physics such as thermalization~\cite{lazarides_periodic_2014,abanin_effective_2015,abanin_exponentially_2015,kuwahara_floquetmagnus_2016,mori_rigorous_2016} and phase structure~\cite{khemani_phase_2016}.

However, relatively little attention has been paid to driven systems at criticality, whose low-energy dynamics are often described by a conformal field theory (CFT). Such quantum critical systems are a natural setting in which to study Floquet dynamics, as many insights into the non-equilibrium dynamics of many-body systems have come from the study of CFTs in 1+1d~\cite{PhysRevLett.96.136801,1742-5468-2005-04-P04010,1742-5468-2007-10-P10004,1742-5468-2016-6-064003}. A na\"ive expectation is that such a driven critical system would simply heat up. However, in the presence of a boundary drive, the energy injected per cycle is not extensive in system size, and there are multiple possible behaviors in an arbitrarily long period prior to thermalization. Moreover, as CFTs are integrable, it is natural to expect they can escape heating even at low frequencies provided the scaling limit is taken before the long time limit. This opens the door to using scaling theory combined with the analytical toolkit of boundary CFT~\cite{cardy_boundary_2006,cardy_conformal_1984,cardy_boundary_1989,Affleck:1996mm} to characterize multiple regimes of universal dynamics in such boundary driven quantum critical points.

In this Letter, we study the dynamics of entanglement entropy $S_l(t)$ and Loschmidt echo ${\cal L}(t) = \abs{\braket{\psi(0)}{\psi(t)}}^2$ in conformally-invariant quantum critical systems subject to a periodic boundary drive. We find two distinct regimes in which boundary conformal field theory provides an excellent description of the dynamics. For suitably slow drives, the system behaves almost as though subject to a series of independent quantum quenches but with amplitude corrections related to multiple-point correlation functions, while for fast drives, the boundary drive can be averaged out, and the system responds as though subject to a single quench at an averaged value of the field. For intermediate driving frequency, we find universal heating which crosses over from a perturbative regime at weak drive to non-perturbative boundary CFT regime at strong drive. The dynamics in all driving regimes are universal and can be described using field-theoretic tools. We numerically confirm that the dynamics remain robust against adding integrability-breaking interactions up to the finite times that may be simulated.

\paragraph{Model.} While our results apply to arbitrary boundary-driven CFTs, for concreteness we will focus on the archetypical transverse-field Ising (TFI) model on the half-line with a time-dependent symmetry-breaking boundary field
\begin{equation}
H = - \sum_{i \geq 0} \left( J  \sigma_i^z \sigma_{i+1}^z + h  \sigma_i^x  +\Gamma \sigma_i^x \sigma_{i+1}^x \right) -  h_b(t) \sigma_0^z, 
\label{eq:TFI}
\end{equation}
with $\Gamma$ an integrability-breaking perturbation and $h \sim J$ tuned to the critical point. This model has a convenient description in terms of free fermions when $\Gamma = 0$, seen by performing a Jordan-Wigner transformation~\cite{sachdev2007quantum}\footnote{See supplemental material, which includes references \cite{peschel_calculation_2003,eisler_evolution_2007,peschel_calculation_2003,eisler_evolution_2007,stephan_local_2011,PhysRevA.72.013604,kennes_universal_2014,Schollwock201196,PhysRevLett.69.2863,di_francesco_conformal_2011,giamarchi_quantum_2004,Blanes2009151,abanin_exponentially_2015}.} and is thus an ideal numerical test-bed for our model-independent analytical arguments. We initially prepare the system in the ground state at fixed boundary field $h_b(t<0)$ then quench on a periodic boundary drive, $h_{b}(t+T) = h_b(t)$, for $t \geq 0$. In equilibrium, the low energy description of this spin chain at criticality is well-understood in terms of gapless left- and right-moving Majorana fields satisfying $\lbrace \eta_{R/L} (x), \eta_{R/L} (y)\rbrace = \delta(x-y)$, with Hamiltonian 
\begin{equation}
H = - \frac{i v }{2} \int_{0}^{\infty} dx \left( \eta_R \partial_x \eta_R  -  \eta_L \partial_x \eta_L\right) - \lambda(t) \sigma_b(0),
\label{eq:TFIFieldTheory}
\end{equation}
where we dropped irrelevant terms. Here, $v$ is a non-universal velocity ($v=2 J$ for $\Gamma=0$)  and $\lambda \propto h_b$. In this Majorana formulation, the boundary spin can be represented as $\sigma_b(0) = i (\eta_R + \eta_L) \gamma$~\cite{doi:10.1142/S0217751X94001552}, where $\gamma^\dagger = \gamma$ is an ancilla Majorana satisfying $\gamma^2=1$ that anticommutes with all fields. In the following, we will assume that the drive is characterized by a single scale $\| h_b(t) \| \sim h_b $ which we take to be much smaller than the single particle bandwidth $h_b \ll \Lambda \equiv 2J=2$ (setting $J=1$), for which field theory is a good equilibrium description. The boundary field is a relevant perturbation with scaling dimension $\Delta=\frac{1}{2} < 1$ in the renormalization group (RG) language, with characteristic time scale $t_b \sim |h_b|^{-\nu_b} $, $\nu_b=(1-\Delta)^{-1}=2$.

There are three energy scales in this problem: the driving frequency $\omega=2\pi/T$, the bandwidth $\Lambda$, and the scale of the boundary perturbation $t_b^{-1} \sim h_b^{\nu_b} \ll \Lambda$. We will now consider various orderings of these scales and argue that essentially all regimes can be understood using a combination of field theory and scaling arguments, even though the drive is continuously injecting energy into the system. While the Hamiltonian~\eqref{eq:TFI} for $\Gamma=0$ can be mapped onto free fermions for numerical convenience~\cite{supmat}, we note that our main conclusions follow from general field theory arguments and therefore continue to hold in the non-integrable case. We emphasize that although we choose to focus on the Ising field theory~\eqref{eq:TFIFieldTheory} as an example, our field-theoretic arguments are model-independent, so our results carry over immediately to any boundary driven CFT, such as a driven quantum impurity problem with $t_b^{-1} \to T_K$, the Kondo temperature.

\begin{figure}
	\includegraphics[width = \columnwidth]{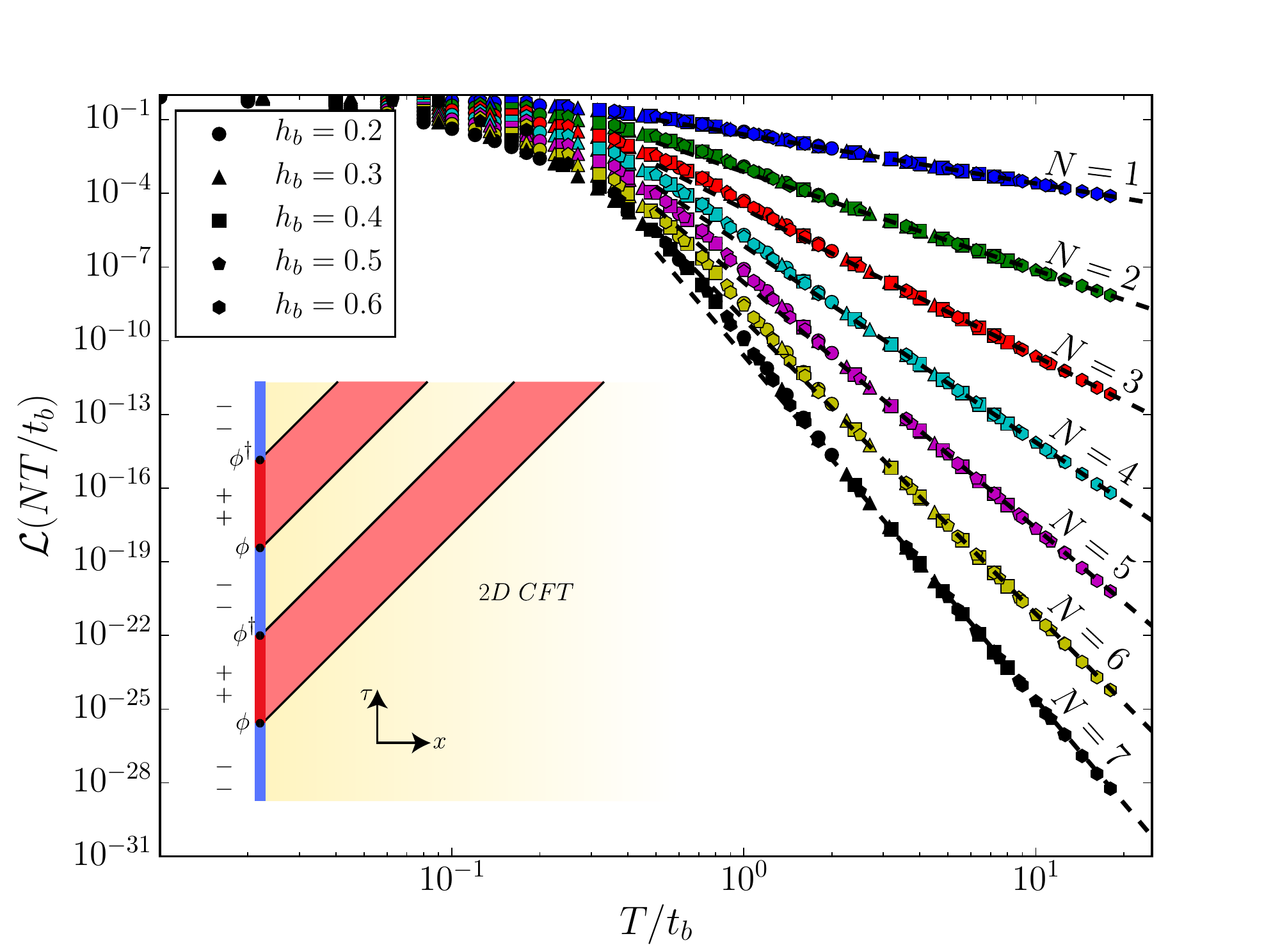}
	\caption{Slow driving regime $\omega \ll t_b^{-1} \sim h_b^2 \ll \Lambda$ for a step drive alternating between $-h_b$ and $+h_b$ for systems up to $L \sim 3200$ sites ($\Gamma=0$). For large $T$, we see clear power-law scaling of the Loschmidt echo with slope $-2 N $ as predicted from boundary CFT. The agreement between the CFT $2N$-point function prediction (dashed lines) and numerical data is excellent, where we stress that the only fit parameter is the non-universal offset $c_1$. Note also the universal collapse of the Loschmidt echo as a function of universal parameter $T / t_b \sim h_b^2 T$. Inset: sketch of the imaginary time picture where the step drive corresponds to inserting BCC operators.}
	\label{figslow}
\end{figure}

\paragraph{Slow driving regime: step drive.} We start by considering the slow driving regime $\omega \ll t_b^{-1} \ll \Lambda$ for a step drive starting from the initial field $h_b(t<0) = -h_b$ with $h_b (t) = +h_b$ for $0 \leq t \leq T/2$ (Hamiltonian $H_1$) and $h_b (t) = -h_b$ if $T/2 \leq t \leq T$  (Hamiltonian $H_0$) for $t\geq 0$. Intuitively, this drive looks like independent local quenches. Focusing on the Loschmidt echo (return probability) ${\cal L} = \left| \langle \psi_0   \ket{\psi(t)}  \right|^2$~\cite{Goussev:2012}, this behavior is best understood by Wick rotating to imaginary time $\tau = i t$, where the spin-chain Loschmidt echo can be mapped onto a CFT correlation function. After computing this correlation function, we Wick rotate back to real time to obtain the dynamical echo. In imaginary time, the initial state can be generated by an infinite imaginary time evolution $\lim_{\tau\to\infty} e^{-\tau H_0}\ket{0} \propto \ket{\psi_0}$ from arbitrary initial state $\ket{0}$. In imaginary time, $\exp(-\tau H)$ acts as a projector onto the ground state of $H$, so for large $T \gg t_b$ we essentially oscillate between the ground states of $H_0$ and $H_1$, for which $\sigma_0^z$ is locked in the direction of the boundary field $\pm h_b$. In the CFT language, a sharp change in boundary conditions can be treated by inserting a boundary-condition changing (BCC) operator~\cite{cardy_boundary_2006}, as diagrammed in Fig.~\ref{figslow}. This means that the Loschmidt echo $\mathcal L(NT)$ after $N$ periods of drive corresponds to the $2N$-point correlation function of a BCC operator $\phi_{\rm BCC}$ changing the boundary condition from fixed $\sigma_0^z= \pm 1$ to $\sigma_0^z= \mp 1$. 

Analytically continuing to real time, we expect the Loschmidt echo to be a universal function ${\cal L} (T /t_b,N)$ in the field theory regime. In the limit $T \gg t_b$, this reduces to the $2N$-point function
\begin{equation}
{\cal L}(NT) \underset{T \gg t_b}{\sim} \abs{\avg{\prod_{n=0}^{2N-1}\phi_{{\rm BCC}}(n T/2)}}^2 = c_N  \left(\frac{T}{t_b}\right)^{- \gamma N} ,
\label{eqLoschQuench}
\end{equation}
whose form is fixed by scale invariance. The universal exponent $\gamma = 4 h_{+-}=2$ is given by the scaling dimension $h_{+-}=\frac{1}{2}$ of the BCC operator $\phi_{\rm BCC}$~\cite{cardy_conformal_1984,cardy_boundary_1989}.  Other step drives can be dealt with in a similar fashion; for example, a step drive from $h_b=0$ to $h_b \neq 0$ corresponds to the insertion of a BCC field with scaling dimension $h_{\rm BCC}=\frac{1}{16}$. We emphasize that eq.~\eqref{eqLoschQuench} holds for arbitrary boundary step drives in more general CFTs with the appropriate choice of BCC operator.

Note that although the Loschmidt echo decays exponentially with $N$, consistent with the independent quenches picture, the fact that the quenches are not fully independent is encoded in the non-trivial $N$ dependence of the coefficients $c_N$. The ratio $c_N/(c_1)^N$ is universal and can be computed exactly for this specific drive, since the BCC operator $\phi_{\rm BCC}$ corresponds to a chiral fermionic field $\psi$ in the Ising field theory with $2N$-point correlator given by a Pfaffian: ${\cal L}(NT) \sim \abs{\avg{\psi(0) \psi(T/2) \psi (T) \dots }}^2 \sim \left| {\rm Pf}( 1/(t_i-t_j)) \right|^2 $ with $t_i=0,T/2,T, \dots, (N-\frac{1}{2})T$. For step drives in general CFTs, such universal ratios can be computed within the Coulomb gas (bosonization) framework~\cite{supmat}.  These analytical expressions are in excellent agreement with numerical simulations for $\Gamma=0$ (Fig.~\ref{figslow}), where the only non-universal fit parameter is $c_1$. Since these predictions rely solely on field theory, they apply equally well to the non-integrable case $\Gamma \neq 0$; the interactions $\Gamma \sigma_i^x \sigma_{i+1}^x$ are irrelevant in the RG sense and therefore do not change the universality class. We confirm this numerically by locating the new critical point for $\Gamma \neq 0$ using exact diagonalization, obtaining the ground state using standard density matrix renormalization group (DMRG) techniques~\cite{PhysRevLett.69.2863,Schollwock201196}, and simulating the dynamics of this driven interacting chain using time-evolving block decimation (TEBD)~\cite{PhysRevLett.93.040502}. We find excellent agreement with our field-theoretic argument, as shown in the Supplemental Material~\cite{supmat}.


\begin{figure}
	\includegraphics[width = \columnwidth]{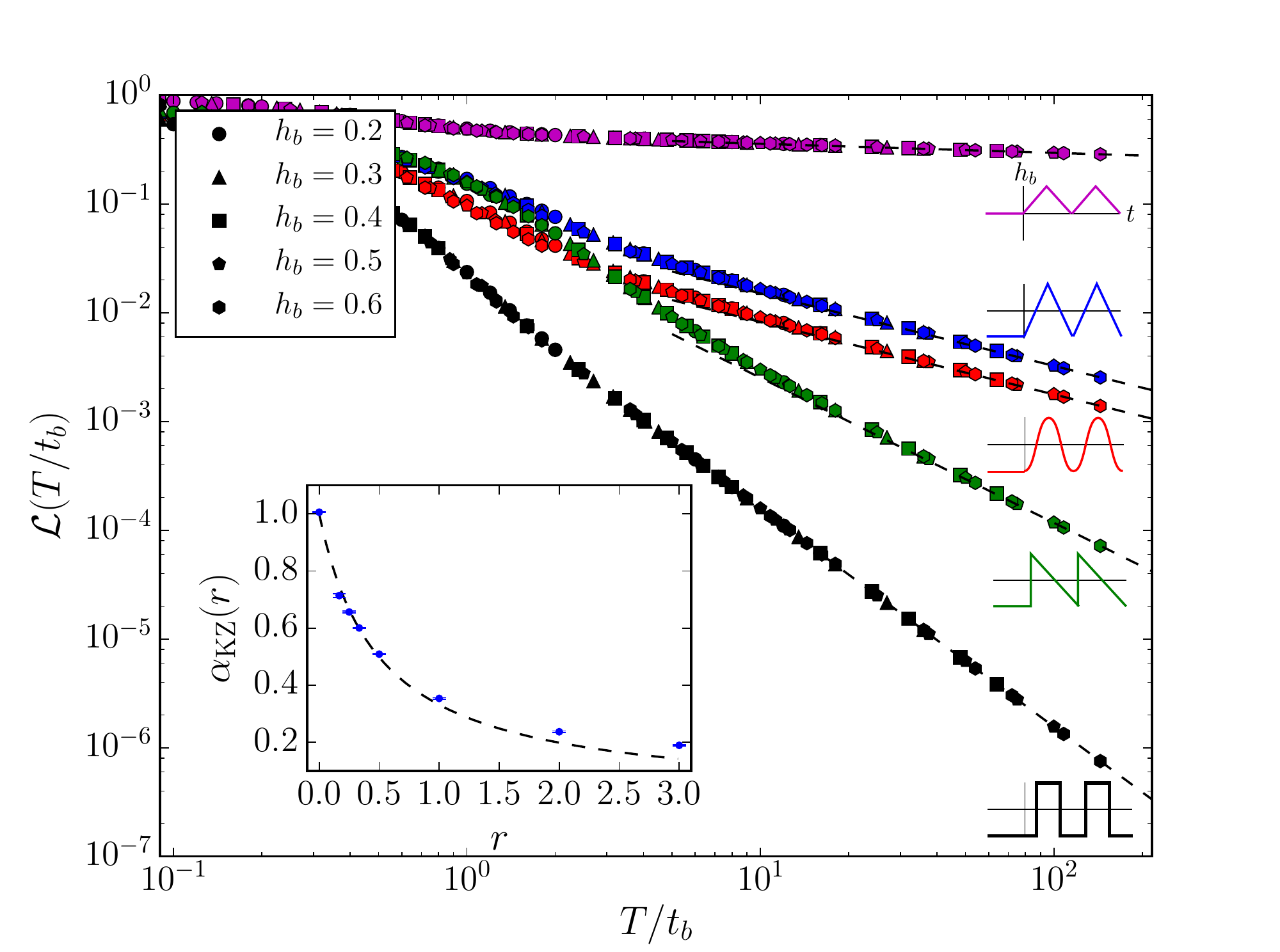}
	\caption{Loschmidt echo for $\Gamma=0$ over a single cycle ($N=1$) in the slow regime for various drive geometries showing renormalized power laws and universal collapse as a function of $T/t_b = T h_b^2$. The dashed lines correspond to the analytic prediction~\eqref{eqKZexp} from boundary CFT and KZ arguments. Inset: KZ renormalization factor $\alpha_{\rm KZ}$ of the BCC exponents for a boundary field scaling as $h_b(t) = h_b  (\frac{t}{T})^r$ compared to the KZ prediction $\alpha_{\rm KZ}= (1+\nu_b r)^{-1}$ with $\nu_b=2$. The dashed line is a fit of the numerical data for small $r$ giving $\nu_b \approx 2.02 \pm 0.08$. }
	\label{FigKZ}
\end{figure}

\textit{Slow driving regime: general drives.} Consider now a more general drive such as $h_b(t>0) = -h_b \cos(\pi t / T)$ with $h_b(t<0)=-h_b$. In the large $T$ limit $h_b(t)$ crosses the critical value slowly rather than suddenly, yet the BCC picture suggests that the field should quickly flow to infinity. We find, however, that the vanishing (but finite) crossing speed is strongly relevant, changing the power law entirely (Fig.~\ref{FigKZ}). To understand this difference, we use the concept of Kibble-Zurek (KZ) scaling, which is frequently applied to bulk drives crossing a bulk quantum critical point~\cite{PhysRevLett.95.105701,PhysRevB.72.161201,PhysRevLett.95.245701,2016arXiv161202259L}  but has not been studied for such boundary drives to our knowledge.

Let us imagine that the drive crosses $h_b=0$ as a power-law $h_b(t) = h_b  |\frac{t}{T}|^r \mathrm{sgn}(t)$ with $r=1$ in the cosine drive considered above and $r = 0$ for a quench~\cite{PhysRevB.81.012303}. The effective time scale $t_b(t) \sim \left[ h_b(t) \right]^{-\nu_b} $ now becomes time-dependent, and we expect the dynamics to be controlled by an emergent time scale
\begin{equation}
t_{\rm KZ} \sim T^{\frac{r \nu_b}{1+r \nu_b}} h_b^{-\frac{ \nu_b}{1+r \nu_b}},
\label{eqKZscale}
\end{equation}
%
%
given by $t_{\rm KZ} \sim t_b(t_{\rm KZ})$. Though our system is always gapless so that there is no adiabatic limit, it is straightforward to show that this dynamical scale emerges directly from the equations of motion of eq.~\eqref{eq:TFIFieldTheory} \cite{mike_prl}. It is natural to expect that the slow driving limit $T \gg t_{\rm KZ}$ should still be described by boundary CFT, suggesting that the Loschmidt echo would scale as~\eqref{eqLoschQuench} with $t_b$ replaced $t_{\rm KZ}$. We therefore see that the effect of the slow driving amounts to renormalizing the dimension $h_{\rm BCC}$ of the BCC operator by a factor $\alpha_{\rm KZ} = 1/(1+r \nu_b)$ with $\nu_b=2$ in our case. More generally, for a drive where $h_b(t)$ crosses or touches the critical value $n$ times within a single cycle, we predict that the universal exponent $\gamma$ controlling the exponential decay of the Loschmidt echo is given by 
\begin{equation}
\gamma = 2 \sum_{i=1}^n \frac{h^i_{\rm BCC}}{1+r_i \nu_b},
\label{eqKZexp}
\end{equation}
where $r_i$ is the power of $|h_b(t)| \sim |t-t^i_c|^{r_i}$ near the critical time $t^i_c$. For our model, $h^i_{\rm BCC} =\frac{1}{2}$  if $h_b(t)$ crosses zero and $h^i_{\rm BCC} =\frac{1}{16}$ if it touches zero without changing sign~\cite{cardy_conformal_1984,cardy_boundary_1989,cardy_boundary_2005}. For example, a cosine or triangle drive oscillating between $\pm h_b$ has $n=2$, $r_1=r_2=1$ so that $\gamma = 2/3$, while a sawtooth drive combines slow ($r_1=1$) and fast ($r_2=0$) crossings to give $\gamma=4/3$. 

These predictions give good agreement with numerics (Fig.~\ref{FigKZ}) \footnote{We note that the agreement is systematically worse for larger $r$ due to finite system size and Trotter step errors.}. Furthermore, the only effect of the slow driving is to renormalize the scaling dimensions of the BCC operators while keeping the structure of the $2N$-point function unchanged. In particular, we find that the universal numbers $c_N/(c_1)^N$ in eq.~\eqref{eqLoschQuench} are still given by the boundary CFT predictions for a step drive~\cite{supmat}.

\begin{figure}
	\includegraphics[width = 0.9\columnwidth]{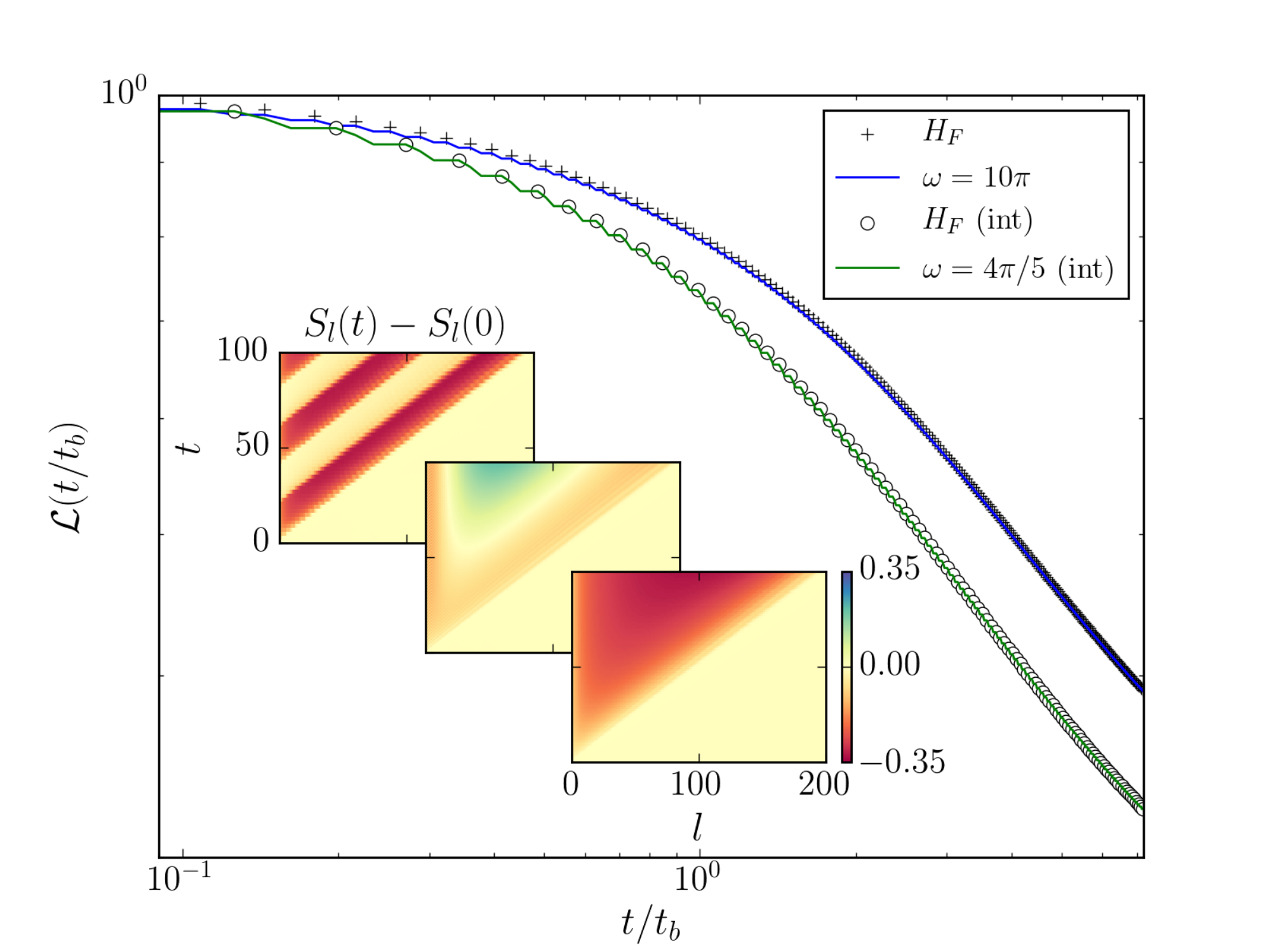}
	\caption{Fast regime: the Loschmit echo at frequencies $\omega > \Lambda$ for a step drive oscillating between $0$ and $h_b$ coincides with the echo after a single local quench with effective field $h_b/2$ with Floquet Hamiltonian $H_F$ (black crosses). This result also holds when interactions are added with $\Gamma=0.25$ (white circles, green line). Insets: entanglement entropy difference $S_l(t) - S_l(0)$ for $\Gamma = 0$ as the drive frequency crosses over from the intermediate to the fast regime.}
	\label{FigFast}
\end{figure}

\paragraph{Fast driving regime.} We now consider the high-frequency regime $t_b^{-1} \ll \Lambda \ll \omega$. This is na\"ively outside the regime where field theory results should apply, but we can take advantage of standard Floquet machinery to write a Floquet-Magnus high-frequency expansion for the Floquet Hamiltonian $H_F$ defined by $U(T) ={\cal T}  {\rm e}^{- i \int_{0}^T dt H(t)}= {\rm e}^{-i T H_F}$~\cite{CPA:CPA3160070404}. For example, $H_F = \frac{1}{2}(H_0 + H_1) - \frac{i }{4 \omega} [H_0,H_1] + \mathcal O (\omega^{-2}) $ for a step drive. While higher order terms in this expansion are suppressed by powers of $\omega^{-1}$ as for any high-frequency Floquet system, we note here that the Floquet Hamiltonian $H_F$ itself corresponds to a CFT subject to an effective boundary field $\overline{h}_b = (1/T) \int_0^T h_b(t) dt$ with higher order terms in the high frequency expansion being RG irrelevant. This is most easily seen using the field theory Hamiltonian~\eqref{eq:TFIFieldTheory} where the small parameter controlling the expansion is $v/\omega \ll 1$ with $v \sim \Lambda = 2J$. While the first boundary term has scaling dimension $\Delta=1/2$ and corresponds to the averaged field $\overline{h}_b$, dimensional analysis immediately implies that terms of order $\omega^{-n}$ have scaling dimension of  at least $n+1/2$ due to terms such as $\partial^{n} \eta (0)$ and are thus irrelevant for $n>0$~\cite{supmat}. Therefore at late times, the system behaves as though subject to a single local quantum quench with effective boundary field $\overline{h}_b$ (Fig.~\ref{FigFast}), a problem whose universal dynamics has been studied extensively~\cite{PhysRevLett.110.240601,PhysRevX.4.041007}. We remark that though RG techniques may be in general ill-defined in a Floquet system which, for instance, lacks a notion of ground state, in this high-frequency limit the Floquet evolution is well-controlled by an effective static Hamiltonian. Since our initial state is a conformally invariant ground state and the effective Hamiltonian implements a local quench, the notion of RG flow is well-defined~\cite{PhysRevLett.110.240601} and provides a powerful tool of analysis. Additionally, for the non-interacting (free fermions) case with $\Gamma=0$ in eq.~\eqref{eq:TFI}, one may prove that the high-frequency expansion is convergent for $\omega \gae \Lambda$ by bounding the spectral width of the single-particle Hamiltonian~\cite{supmat}. More generally, this effective single quench picture will survive even in the presence of integrability-breaking interactions controlled by $\Gamma$ up to exponentially long time scales $\tau_{\rm th} \sim {\rm e}^{C \omega/ \Lambda}$ \cite{abanin_effective_2015,abanin_exponentially_2015,mori_rigorous_2016,kuwahara_floquetmagnus_2016}. We simulated the dynamics of this interacting chain subject to the same drive using TEBD and found excellent agreement with the single effective quench picture even at moderate frequencies (Fig.~\ref{FigFast}).


\begin{figure}
	\includegraphics[width = \columnwidth]{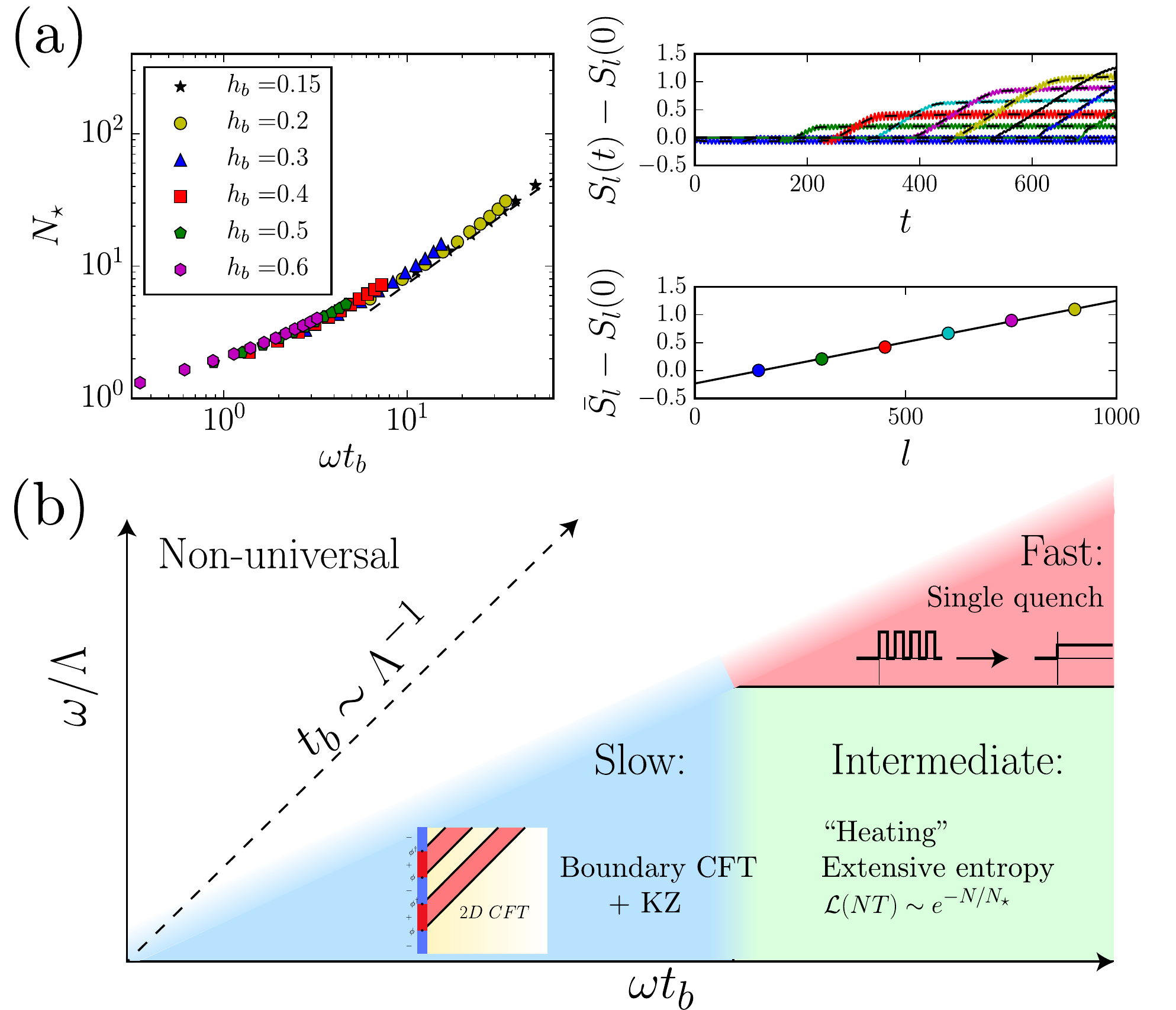}
	\caption{Intermediate regime $t_b^{-1} \sim \omega \ll \Lambda $ for a step drive from $h_b = 0$ to $h_b \not=0$. (a) Left panel: universal scaling function $N_\star( \omega t_b)$ characterizing the exponential decay of the Loschmidt echo ${\cal L} (N T) \sim {\rm e}^{-N/N_\star}$, with the dashed line showing the linear behavior expected from Fermi's golden rule. Right panel: volume-law scaling of the entanglement entropy in the long-time limit for $\omega t_b = 10.5 \gg 1$. An overbar denotes the value of the late-time plateau. (b) Sketch of the three universal driving regimes analyzed in this Letter.}
	\label{FigIntermediate}
\end{figure}

\textit{Crossover regime.} Finally, we discuss the intermediate crossover regime $t_b^{-1} \sim \omega \ll \Lambda$. We focus on a free-to-fixed step drive from $h_b = 0$ to $h_b \not=0$ with $\Gamma=0$ for simplicity. In this regime, we expect the system to absorb energy (``heat'') via resonant processes within the single-particle bandwidth. This leads to exponential decay of the Loschmidt echo,
\begin{equation}
{\cal L} (N T) \underset{t_b^{-1}, \omega \ll \Lambda} {\sim}{\rm e}^{-N/N_\star( \omega t_b)},
\end{equation}
with $N_\star( \omega t_b)$ a universal function (Fig.~\ref{FigIntermediate}a). For weak drive ($\omega t_b \gg 1$), resonant heating occurs with a rate $\tau^{-1} \sim h_b^2/J $ given by Fermi's golden rule, so that  $N_\star \sim \tau /T \sim \omega t_b$. For strong drive ($\omega t_b \ll 1$), we recover the boundary CFT prediction $N_\star \sim -1/(\gamma \log \omega t_b)$. We also find that entanglement entropy of boundary intervals of size $\ell$, relative to the ground state entropy, saturates to a volume law behavior $S_\ell \sim \ell$ at long times in the regime $ \omega t_b \gg 1$, consistent with heating~\footnote{We note that by heating we simply mean absorption of energy, not thermalization. Since this is an integrable model, the reduced density matrix should more accurately be described by a generalized Gibbs ensemble~\cite{PhysRevLett.98.050405} which in some observables can look highly athermal}. At low frequencies, the entropy simply oscillates between ground state values~\footnote{In the absence of boundary field, at large distances the ground state entanglement entropy is given by the well-known result $S_l(h_b=0)=\frac{c}{3} \ln \frac{l}{\tilde a}$\cite{1751-8121-42-50-504005}, where $c=1/2$ is the central charge of the CFT and $\tilde a$ is a non-universal constant. Pinning the boundary spin causes a universal reduction of the entanglement entropy for $h_b \neq 0$ of $\Delta S = -\log \sqrt 2$, known as the Affleck-Ludwig entropy \cite{affleck_universal_1991}.} though it may become extensive at much later times. We leave a detailed analysis of the role of interactions in this intermediate regime for future work. 

\paragraph{Discussion.} We have investigated CFTs subject to a Floquet boundary drive. Despite the na\"ive expectation that such gapless systems should absorb energy and simply heat up, we have identified three distinct regimes summarized in Fig.~\ref{FigIntermediate}b in which the system shows universal features that can be understood using tools of field theory and scaling theory. We expect our main conclusions to apply to a broad class of systems, and it will be especially interesting to investigate the consequences of our results for the physics of driven quantum dots and the non-equilibrium signatures of topological edge modes~\cite{jiang_majorana_2011}. In general, our results represent an analytically tractable model of a driven gapless system, an active area of research increasingly relevant to experiments.


\paragraph{Acknowledgments.} We thank J. Cardy, R. Ilan, A. Polkovnikov, and W.-W. Ho for useful discussions. This work used the Extreme Science and Engineering Discovery Environment (XSEDE)\cite{john_towns_xsede:_2014}, which is supported by National Science Foundation grant number ACI-1053575. W.B. acknowledges support from the Department of Defense (DoD) through the National Defense Science \& Engineering Graduate Fellowship (NDSEG) Program, and from the Hellman Foundation through a Hellman Graduate Fellowship. We also acknowledge support from Laboratory directed Research and Development (LDRD) funding from Berkeley Laboratory, provided by the Director, Office of Science, of the U.S. Department of Energy under Contract No. DEAC02-05CH11231 (M.K. and R.V.), from the U.S. DOE, Office of Science, Basic Energy Sciences (BES) as part of the TIMES initiative (M.K.), from NSF DMR-1507141 and a Simons Investigatorship (J.E.M.), and center support from CaIQuE and the Moore Foundation's EPiQS initiative. M.K., R.V., and J.E.M. acknowledge the hospitality of KITP, supported in part by the National Science Foundation under Grant No. NSF PHY-1125915.

\bibliography{FloquetCFT}
\bibliographystyle{apsrev4-1}


\bigskip

\onecolumngrid

\newpage



\section*{Supplementary material}

# Floquet Dynamics of Boundary-Driven Systems at Criticality: Supplemental Material

William Berdanier*
*Department of Physics, University of California, Berkeley, CA 94720, USA*

Michael Kolodrubetz, Romain Vasseur, and Joel E. Moore
*Department of Physics, University of California, Berkeley, CA 94720, USA and*
*Materials Sciences Division, Lawrence Berkeley National Laboratory, Berkeley, CA 94720, USA*
(Dated: May 23, 2017)

## I. NUMERICAL METHODS: FREE FERMIONS AND INTERACTIONS

In this appendix, we provide details on the numerical simulation of the transverse-field Ising (TFI) chain with longitudinal field, and the extraction of the entanglement entropy and the Loschmidt echo. To numerically study the time evolution of the TFI chain ground state subject to the Floquet drive described above, we utilize the fact that the non-interacting TFI chain can be efficiently described as a system of free Majorana fermions. This fact will prove useful in calculating the entanglement entropy of subsections of the chain as a function of time [1, 2]. We cover the free case first, then move on to the interacting one in subsection C. For completeness, the TFI Hamiltonian in terms of spins is

$$H = -\sum_{j=0}^{L-1}(J\sigma_i^z\sigma_{i+1}^z + h\sigma_i^x + \Gamma\sigma_i^x\sigma_{i+1}^x) - h_b(t)\sigma_0^z \quad (1)$$

with interactions controlled by $\Gamma$.

### A. Entanglement entropy

First, let us apply a Jordan-Wigner transformation to the non-interacting TFI chain $\sigma_j^x = i\gamma_{B,j}\gamma_{A,j}$, $\sigma_j^z = -i\gamma\left(\prod_{l=0}^{j-1} i\gamma_{A,l}\gamma_{B,l}\right)\gamma_{B,j}$ where the $\gamma$ operators obey the Majorana algebra $\{\gamma_{\alpha,i}, \gamma_{\beta,j}\} = 2\delta_{\alpha\beta}\delta_{ij}$, $\gamma_{\alpha,i}^2 = 1$. We include in our Jordan-Wigner transformation an ancilla Majorana operator $\gamma$ that plays no dynamical role. In the Majorana language, then, the TFI Hamiltonian is

$$H = \sum_{j=0}^{L-2} i\gamma_{A,j}\gamma_{B,j+1} + \sum_{j=0}^{L-1} i\gamma_{A,j}\gamma_{B,j} + h_0 i\gamma\gamma_{B,0}. \quad (2)$$

at the critical point $h = J = 1$. This doubles the size of the Hilbert space, making each original level doubly degenerate. Now, if we are in a state satisfying Wick's theorem, everything is essentially determined by the two-point correlator $C_{ij} = \langle\gamma_i\gamma_j\rangle$. The Majorana anticommutation relation implies that $C_{ij} = 2\delta_{ij} - C_{ji}$, so $C_{ij} = \delta_{ij} + a_{ij}$, where $a$ is some antisymmetric matrix. Let us now diagonalize $a = q^T\sigma q$, where $q$ is orthogonal and $\sigma$ has form $\sigma = \text{diag}\begin{pmatrix} 0 & \lambda_i \\ -\lambda_i & 0 \end{pmatrix}_{i=1}^L$. This form has the eigenvalues arranged such that $\sigma_{i,i+1} = \lambda_i$, and satisfies $\sigma^T = -\sigma$. Now define $\gamma' = q\gamma$. Then

$$\langle \gamma'_{2k'-1}\gamma'_{2k}\rangle = q_{2k'-1,i}q_{2k,j}(\delta_{ij} + \lambda_{k''}(q_{2k''-1,i}q_{2k'',j} - q_{2k'',i}q_{2k'',j})). \quad (3)$$

From the orthogonality of $q$, $q_{\alpha i}q_{\beta i} = \delta_{\alpha\beta}$, so the only non-vanishing term is $q_{2k'-1,i}q_{2k,j}\lambda_{k''}q_{2k''-1,i}q_{2k'',j} = \lambda_k\delta_{kk'}\delta_{k'k''}$. Thus the only non-vanishing two-point function is

$$\langle \gamma'_{2k-1}\gamma'_{2k}\rangle = -\langle \gamma'_{2k}\gamma'_{2k-1}\rangle = \lambda_k. \quad (4)$$

We can write this correlation function as arising from a single particle density matrix $\rho = \frac{1}{Z}\prod_k e^{i\epsilon_k\gamma'_{2k-1}\gamma'_{2k}}$. Now, we can construct a complex fermionic operator from Majorana operators via $c_k = \frac{\gamma'_{2k-1}+i\gamma'_{2k}}{2}$. This gives $\gamma'_{2k-1}\gamma'_{2k} = -i(2c_k^\dagger c_k - 1)$. This gives the density matrix as

$$\rho = \prod_k \frac{e^{\epsilon_k(2n_k-1)}}{e^{\epsilon_k} + e^{-\epsilon_k}}.$$

Thus the two-point function is $\langle \gamma'_{2k-1}\gamma_{2k}\rangle = -i\langle 2n_k - 1\rangle = \lambda_k = -i\frac{e^{\epsilon_k}(+1) + e^{-\epsilon_k}(-1)}{e^{\epsilon_k} + e^{-\epsilon_k}} = -i\tanh\epsilon_k$. Thus we have the non-trivial relation $i\lambda_k = \tanh\epsilon_k$. Now define $\mu_k = |\lambda_k|$. Then $\epsilon_k = \tanh^{-1}(\mu_k)$. To find the entanglement entropy, write the density matrix as

$$\rho = \prod_k \left[p_k|0_k\rangle\langle 0_k| + (1-p_k)|1_k\rangle\langle 1_k|\right], \quad p_k = \frac{e^{-\epsilon_k}}{e^{\epsilon_k} + e^{-\epsilon_k}}. \tag{5}$$

Then the entanglement entropy is $S = -\text{Tr}\rho\log\rho = -\sum_{k=1}^L p_k\log p_k + (1-p_k)\log(1-p_k)$. To time evolve the correlation function, in the Heisenberg picture $C_{ij}(t) = \langle\gamma_i(t)\gamma_j(t)\rangle$. Now, for $A$ and $B$ Majorana operators, $e^{\alpha AB} = \cos\alpha + AB\sin\alpha$. Thus, $e^{\alpha AB}Ae^{-\alpha AB} = A\cos 2\alpha - B\sin 2\alpha$, and $e^{\alpha AB}Be^{-\alpha AB} = B\cos 2\alpha + A\sin 2\alpha$. Thus, defining the diagonal matrix

$$D(t) = \text{diag}\left[\begin{pmatrix}\cos(2\epsilon_k t) & -\sin(2\epsilon_k t) \\ \sin(2\epsilon_k t) & \cos(2\epsilon_k t)\end{pmatrix}\right]_{k=1}^{2L} \tag{6}$$

we find that $\gamma'_i(t) = D_{ij}(t)\gamma'_j(0)$. Defining $\Gamma(t) \equiv Q^T D(t)Q$, we see that the correlation function evolves particularly simply as $C(t) = \Gamma(t)C(0)\Gamma(t)^T$. From the time-evolved correlation function we can dynamically calculate the entanglement entropy as above.

### B. Loschmidt echo

Calculation of the Loschmidt echo directly from the Majorana operator two-point function above is a bit more challenging due to the fact that the TFI Hamiltonian does not conserve particle number when written in terms of complex fermion operators. This is not an issue for a single quench, as has been explored in Refs. 1–3 and many others, but becomes quite complicated even for three quenches. Instead, it is convenient to use the fact that the $XX$ model – which does conserve particle number – can be decomposed into two independent copies of the TFI chain. The mapping proceeds as follows. Take the $XX$ chain with an ancilla fermion at "site" 0:

$$H_{XX} = -J\sum_{i=1}^{L}(\sigma_i^x\sigma_{i+1}^x + \sigma_i^y\sigma_{i+1}^y) - J'\sigma_0^x\sigma_1^x. \tag{7}$$

The total length of the chain is $L+1$ in this notation. Via a Jordan-Wigner transformation, namely $c_i^\dagger = \sigma_i^+\prod_{j<i}\sigma_j^z$, $c_i = \left(\prod_{j<i}\sigma_j^z\right)\sigma_i^-$ with $\sigma^\pm = (\sigma^x \pm i\sigma^y)/2$, we get that the $XX$ Hamiltonian is $H_{XX} = -\frac{J}{2}\sum_{i=1}^L c_i^\dagger c_{i+1} - \frac{J'}{2}c_0^\dagger c_1 + h.c.$ Now we can decompose each fermion operator into two Majorana operators, via $c_i^\dagger = (\gamma_{A,i} - i\gamma_{B,i})/2$, $c_i = (\gamma_{A,i} + i\gamma_{B,i})/2$, where $\gamma^2 = 1$, $\gamma^\dagger = \gamma$ and $\{\gamma_{\alpha,i},\gamma_{\beta,j}\} = 2\delta_{ij}\delta_{\alpha\beta}$. Thus,

$$H_{XX} = -\frac{J}{4}\sum_{i=1}^L i(\gamma_{A,i}\gamma_{B,i+1}) - \frac{J'}{4}i\gamma_{A,0}\gamma_{B,1} - \frac{J}{4}\sum_{i=1}^L i(\gamma_{A,i+1}\gamma_{B,i}) - \frac{J'}{4}i\gamma_{A,1}\gamma_{B,0}. \tag{8}$$

We now see that the $XX$ chain is simply two uncoupled copies of the TFI chain. Thus, the Loschmidt echoes will be related by $\mathcal{L}_{XX}(t) = \mathcal{L}_{TFI}(t)^2$. To get the Loscmidt echo of the $XX$ chain with a driven first link, we generalize the methods of Refs. 4 and 5 to handle multiple quenches between two $XX$ Hamiltonians $H_0$ and $H_1$. We first write the ground state as a filled Fermi sea,

$$|\psi(0)\rangle = \prod_{m=1}^{N_f}\left(\sum_{j=1}^L P_{jm}c_j^\dagger\right)|0\rangle \tag{9}$$

where $N_f$ is the total number of negative eigenvalues of the initial Hamiltonian $H_0$ and the $L \times N_f$ matrix $P$ is the (sorted) matrix of corresponding eigenvectors. Now, under time evolution, $|\psi(t)\rangle = \ldots e^{-iH_0 T/2}e^{-iH_1 T/2}|\psi(0)\rangle = \hat{U}(t)|\psi(0)\rangle$, so $|\psi(t)\rangle = \prod_{m=1}^{N_f}\left(\sum_{j=1}^L P_{jm}(t)c_j^\dagger\right)|0\rangle$ with $P(t) = U(t)P$ where $U(t)$ is an $L \times L$ matrix. This

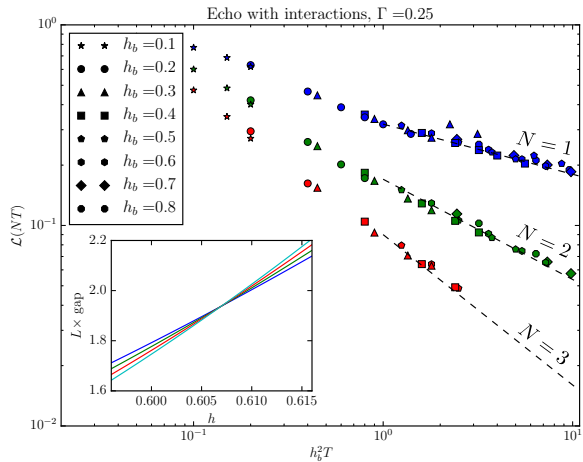

FIG. 1. Loschmidt echo as a function of time in the interacting case, with $\Gamma = 0.25$, for a step drive from $h_b = 0$ to $h_b \neq 0$. Arguments in the main text show that the echo should decay as $\mathcal{L}(NT) \sim T^{-N/4}$ and should be a universal function of $h_b^2 T$, in good agreement with TEBD simulations on $L = 400$ sites. Inset: determination of the new critical point with interactions from exact diagonalization on systems of size $L = 18, 20, 22, 24$. For $\Gamma = 0.25$, the new critical point is at $h = 0.6066(2)$ with $J = 1$.

straightforwardly gives the Loschmidt echo as

$$\begin{aligned}
\mathcal{L}_{XX}(t) &= \left|\langle\psi(0)|\ldots e^{-iH_0 T/2}e^{-iH_1 T/2}|\psi(0)\rangle\right|^2 \\
&= \left|\langle 0| \prod_{m=1}^{N_f}\sum_{j=1}^{L} P_{jm}^* c_j \prod_{n=1}^{N_f}\sum_{i=1}^{L} P_{in}(t) c_i^\dagger |0\rangle\right|^2 \\
&= \left|\det(P^\dagger P(t))\right|^2
\end{aligned} \quad (10)$$

We then simply compute $\mathcal{L}_{TFI}(t) = \sqrt{\mathcal{L}_{XX}(t)}$ to get the desired echo.

### C. Interactions

In the presence of nonzero $\Gamma$, the integrability of the TFI chain is broken, and a description in terms of free fermion operators no longer holds, as the $\sigma_i^x \sigma_{i+1}^x$ terms produce four-fermion interaction terms after a Jordan-Wigner transformation. While interactions break integrability and ultimately lead to late-time thermalization among other effects, they crucially have no effect on the underlying CFT because they are irrelevant in the renormalization-group (RG) sense. We therefore expect the late-time Loschmidt echo to display the same power law behavior as in the free case, albeit with minor deviations at finite times due to the presence of an irrelevant operator. We note that this is true in the appropriate scaling limit $h_b \to 0$ with $h_b^2 T$ fixed and large; for finite $h_b$, as $T \to \infty$ the effects of backscattering would lead to a "dangerous irrelevance" of the interactions that could significantly alter the physics.

We first note that interactions will shift the critical point away from the self-dual point $h = J = 1$. We determine the location of this new critical point numerically by exact diagonalization by looking at the scaling of the gap for systems sizes $L = 18, \ldots, 24$. We then simulate the dynamics using matrix product states (MPS) techniques [6]. The initial state is determined using standard density matrix renormalization group (DMRG) methods [6, 7], and the Floquet dynamics is simulated using the time-evolving block decimation (TEBD) algorithm based on a fourth-order Trotter decomposition with Trotter time step $dt = 0.2$. We adapt the bond dimension of the MPS in order to keep the discarded weight below $10^{-8}$ throughout the whole time evolution. The Loschmidt echo for a step drive from $h_b = 0$ to $h_b \neq 0$ of an interacting Ising chain $\Gamma = 0.25$ with $L = 400$ sites is shown in Fig. 1. Despite the shorter time scales and smaller system sizes accessible using TEBD compared to the non-interacting case, we find that our field theory predictions agree well with TEBD simulations in the low-frequency regime, confirming the universality of our results.

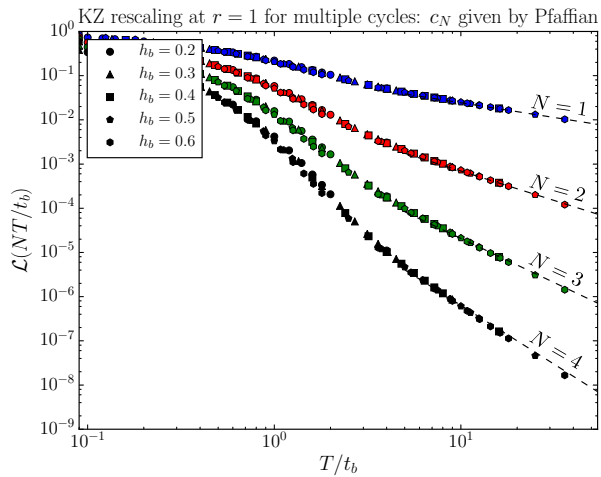

FIG. 2. The boundary Kibble-Zurek scaling mechanism does not affect the $N$-point function structure of the Loschmidt echo. Here we consider a triangle drive from minus to plus, so the power law exponent is rescaled by a factor $1/3$ due to crossing $h_b = 0$ with $r = 1$. The $c_N$ coefficients are still given by a Pfaffian, which is due to the fact that the underlying BCC operator is a chiral fermion with $h_{\rm BCC} = 1/2$.

## II. STRUCTURE OF THE $N$-POINT CORRELATION FUNCTIONS AND KIBBLE-ZUREK SCALING

In the main text we argued that for a step-drive in the low frequency limit, the system is essentially subject to almost independent local quenches, implying a simple exponential decay for the Loschmidt echo (return probability)

$$\mathcal{L}(NT) \underset{T \gg t_b}{\sim} \left| \left\langle \prod_{n=0}^{2N-1} \phi_{\rm BCC}(nT/2) \right\rangle \right|^2 = c_N \left( \frac{T}{t_b} \right)^{-\gamma N}, \tag{11}$$

given by a $2N$-point boundary CFT correlation function. While dependence on the period $T$ is completely fixed by scale invariance with the exponent $\gamma$ being related to the scaling dimension of the operator $\phi_{\rm BCC}$, the behavior of the coefficients $c_N$ is a bit more subtle. In particular, we note that $c_N \neq (c_1)^N$, indicating that the successive local quenches are of course not exactly independent. In fact, the coefficients $c_N/(c_1)^N$ are universal numbers that can be computed using CFT techniques. In the case of a step drive between positive and negative boundary field in the Ising model, this is especially simple since the BCC operator happens to be a fermionic field $\psi$ with dimension $h_{\rm BCC} = \frac{1}{2}$ in the Ising CFT. The $2N$ point function is therefore given by a Pfaffian of an antisymmetric Toeplitz matrix $\mathcal{L}(NT) \sim |\langle \psi(0)\psi(T/2)\psi(T)\ldots\rangle|^2 \sim |{\rm Pf}(1/(t_i - t_j))|^2$ with $t_i = 0, T/2, T, \ldots, (N - \frac{1}{2})T$. This is consistent with $\gamma = 2$, and yields

$$\frac{c_N}{(c_1)^N} = \det \begin{pmatrix} 0 & 1 & \frac{1}{2} & \ddots & \frac{1}{2N-1} \\ -1 & 0 & 1 & \frac{1}{2} & \ddots \\ -\frac{1}{2} & -1 & 0 & 1 & \ddots \\ \ddots & \ddots & \ddots & \ddots & \ddots \\ -\frac{1}{2N-1} & \ddots & \ddots & \ddots & 0 \end{pmatrix}. \tag{12}$$

We emphasize that the normalization with the (non-universal) coefficient of the two-point function $c_1$ is necessary to compare to numerical results.

For more general CFTs and arbitrary step drives, these coefficients can be computed using the Coulomb Gas formalism [8], or in some cases using bosonization. As a simple example, we consider a step drive oscillating between 0 and $h_b$ for which the Loschmidt echo is given by the (chiral) $2N$-point function of the spin operator $\sigma$ with conformal weight $h = \frac{1}{16}$ in the Ising CFT. In order to compute this correlation function, one can "double" the Ising CFT to obtain a free boson theory with central charge $c = 1 = \frac{1}{2} + \frac{1}{2}$ (we already used this trick in Sec. I by expressing the $XX$ chain as two independent copies of the Ising model), and compute the (square of the) $2N$-point function of the spin operator as simple free boson correlator [8]. In principle, more complicated multi-point correlators can also be computed using the Coulomb gas formalism [8].

Remarkably, we note that the boundary Kibble-Zurek (KZ) scaling mechanism at play for more complicated (non-step) drives does not seem to affect the $2N$-point function structure of the Loschmidt echo (Fig. 2). In other words, while the BCC scaling dimensions are renormalized by the KZ mechanism as explained in the main text, the universal ratios $c_N/(c_1)^N$ are still given by the CFT expressions above.

## III. RG ANALYSIS OF THE HIGH FREQUENCY EXPANSION

In this appendix, we detail the high-frequency expansion and renormalization-group (RG) argument described in the main text. We use the so-called Floquet-Magnus (FM) high frequency expansion, a perturbative scheme in the driving period $T$ to compute the Floquet Hamiltonian $H_F[t_0]$ defined from Floquet's theorem by $U(t_0 + T, t_0) = \exp(-iH_F[t_0]T)$. For simplicity we will consider a step drive, though as we shall see a simple scaling argument ensures that our results hold in general. Say that we initially prepare a quantum system in the ground state of some Hamiltonian $H_0$, at time $t = 0$ quench on a different Hamiltonian $H_1$, and at $t = T/2$ again apply $H_0$. Then the Floquet-Magnus high-frequency expansion, fixing the Floquet gauge $t_0 = 0$, takes a particularly simple form, and is in fact just equivalent to the Baker-Campbell-Hausdorff series:

$$H_F = \frac{1}{2}(H_0 + H_1) - \frac{iT}{4}[H_0, H_1] + \frac{(-iT)^2}{24}([H_0, [H_0, H_1]] + [H_1, [H_1, H_0]]) - \frac{(-iT)^3}{48}[H_1, [H_0, [H_0, [H_1]]]] + \mathcal{O}(T^4). \tag{13}$$

Applying this expansion to the transverse-field Ising (TFI) spin chain described in the main text is straightforward; however, more insight can be gleaned from applying the FM expansion directly to the Ising conformal field theory (CFT) itself. Let us first apply an "unfolding" procedure to the Ising CFT [9]. In order to halve the number of fields, we remap $x$ from the half-line to the whole line. We define a new chiral (say, without loss of generality, right-moving) field $\eta(x) = \eta_R(x)$ for $x \geq 0$ and $\eta(x) = \eta_L(-x)$ for $x < 0$ with anticommutation relation $\{\eta(x), \eta(y)\} = \delta(x - y)$. This gives the unfolded Hamiltonian as

$$H = -iv \int_{\mathbb{R}} dx \, \eta(x)\partial_x \eta(x) - i\lambda(t)\gamma\eta(0), \tag{14}$$

with $v = \Lambda = 2J$ and $\lambda \propto h_b$ from the lattice model in the main text, and $\gamma = \gamma^\dagger$ an ancilla Majorana fermion with $\gamma^2 = 1$ that anticommutes with all fields. $H_0$ is then the above Hamiltonian with $\lambda = 0$, and $H_1$ with $\lambda \neq 0$ but constant. At first order, then, $[H_0, H_1] = -\lambda a \int_{\mathbb{R}} dx \, (\eta\{\partial_x \eta, \eta(0)\} - \{\eta, \eta(0)\}\partial_x \eta) = 2\lambda\gamma \int_{\mathbb{R}} dx \, \delta(x)\eta' = 2\lambda\gamma\eta'(0)$. We first note that $[\lambda] = 1 - \Delta = 1/\nu_b = 1/2$ in our case, and $[\eta] = 1/2$ and $[\gamma] = 0$. We can now confirm that the RG dimension of this term is $[\gamma\partial_x \eta] = 3/2 > 1$, so it is RG-irrelevant as claimed in the main text. The second order commutators are likewise

$$[H_0, [H_0, H_1]] = 4i\lambda\gamma\eta''(0), \tag{15}$$

$$[H_1, [H_1, H_0]] = -4i\lambda^2\eta(0)\eta'(0) - i\lambda^2\delta'(0), \tag{16}$$

where we have used the fact that $\{\eta(x), \eta'(y)\} = \partial_x\delta(x - y)$. These operators have RG dimension $[\gamma\partial_x^2\eta] = 5/2$ and $[\eta\partial_x\eta] = [\partial_x\delta] = 2$, so we see that all terms but the lowest order term in the FM high-frequency expansion are irrelevant, getting progressively more irrelevant with higher order in $T$.

This is in fact a feature of any high-frequency expansion applied to this field theory. The zeroth order term will always be just the time-averaged Hamiltonian $\frac{1}{T}\int_0^T dt \, H(t)$ which has RG dimension 1, so factoring out a $\lambda$, the operator itself has dimension $1/2$ and is thus relevant. Now, the $n$th order term, for $n \geq 1$, in any high-frequency expansion will be of the form $\lambda^m T^n \hat{\mathcal{O}}$ with $1 \leq m \leq n$ and $\hat{\mathcal{O}}$ some operator. Since this term has to have units of energy, its overall RG dimension must be 1, so we find

$$[\hat{\mathcal{O}}] = 1 - (1 - \Delta)m + n \geq 1 + \Delta n > 1. \tag{17}$$

In any unitary CFT, $0 < \Delta < 1$ for any relevant perturbation [8], and here $\Delta = 1/2$. Thus, any higher order term must always be irrelevant, regardless of the particular drive and particular expansion considered.

## IV. CONVERGENCE OF THE HFE AND HEATING IN THE HIGH-FREQUENCY LIMIT

In this appendix, we address a couple of questions about the boundary-driven CFT, namely: 1) for the integrable model with $\Gamma = 0$, can we prove convergence of the HFE/Magnus expansion for some region of phase space and 2)

for the CFT with interactions at $\omega > \Lambda$, is there are regime where the physics is described by the CFT quench results before heating takes over? We will argue that the answer to both is yes.

First, consider the integrable, non-interacting TFI model with $\Gamma = 0$. If $C(t) \equiv \langle \eta(t)\eta(t)^T \rangle$ is the Majorana correlation matrix, then we showed above (see Sec. I) that

$$C(t) = U(t)C(0)U(t)^T$$

$$U(t) = \begin{cases} Q_1^T D_1(t) Q_1, & 0 < t < T/2 \\ Q_2^T D_2(t - T/2) Q_2 Q_1^T D_1(T/2) Q_1 & T/2 < t < T \end{cases}$$

$$D_{1,2}(t) = \begin{pmatrix} \cos(2\epsilon_1 t) & \sin(2\epsilon_1 t) & & & \\ -\sin(2\epsilon_1 t) & \cos(2\epsilon_1 t) & & & \\ & & \cos(2\epsilon_2 t) & \sin(2\epsilon_2 t) & \\ & & -\sin(2\epsilon_2 t) & \cos(2\epsilon_2 t) & \\ & & & & \ddots \end{pmatrix},$$

where $Q_{1,2}$ are fixed orthogonal matrices that diagonalize the single-particle Hamiltonian with (positive) eigenenergies $\epsilon_j$. Note that the single-particle bandwidth is $\Lambda = 2\max \epsilon$. Since all physical properties depend solely on the correlation matrix via Wick's theorem, we are interested in doing the Magnus expansion on the single-particle unitary $U(T) = e^{-iH_F T}$. To prove convergence, we will use Theorem 9 of Ref. 10, which states that if $Y' = A(t)Y$ with $Y(0) = 1$, then the Magnus expansion converges if $\int_0^T ||A(s)||ds < \pi$ where $||\cdot||$ denotes the 2-norm. We therefore must massage the above expression into $Y' = AY$. To do so, let us first consider $0 < t < T/2$. Then

$$U'(t) = Q_1^T D_1'(t) Q_1$$
$$= Q_1^T D_1'(t) Q_1 U^T(t) U(t)$$
$$= \underbrace{Q_1^T D_1'(t) D_1(t) Q_1}_{A(t)} U(t),$$

$$D_1' D_1 = \begin{pmatrix} -2\epsilon_1 \sin(2\epsilon_1 t) & 2\epsilon_1 \cos(2\epsilon_1 t) & \\ -2\epsilon_1 \cos(2\epsilon_1 t) & -2\epsilon_1 \sin(2\epsilon_1 t) & \\ & & \ddots \end{pmatrix} \begin{pmatrix} \cos(2\epsilon_1 t) & \sin(2\epsilon_1 t) & \\ -\sin(2\epsilon_1 t) & \cos(2\epsilon_1 t) & \\ & & \ddots \end{pmatrix}$$

$$= \begin{pmatrix} -\epsilon_1 \sin(4\epsilon_1 t) & \epsilon_1 \cos(4\epsilon_1 t) & \\ -\epsilon_1 \cos(4\epsilon_1 t) & -\epsilon_1 \sin(4\epsilon_1 t) & \\ & & \ddots \end{pmatrix}.$$

so that

$$||A|| = ||D_1' D_1|| = \max_i \epsilon_i |\sin(4\epsilon_i t) \pm \cos(4\epsilon_i t)|$$
$$\leq \sqrt{2} \max_i \epsilon_i = \frac{\Lambda}{\sqrt{2}}.$$

So convergence is guaranteed if

$$\int_0^T ||A(s)||ds \leq \frac{T\Lambda}{\sqrt{2}} < \pi,$$

that is, $\Lambda < \frac{\omega}{\sqrt{2}}$. Thus we can prove that for frequencies slightly above the single-particle bandwidth, convergence is guaranteed. Note that this bound is not tight in general, so this is still consistent with the possibility of convergence all the way down to $\Lambda = \omega$. Note also that, up to prefactors, similar proofs should hold for all non-interacting CFTs, meaning that our high-frequency regime is a well-defined dynamical phase.

The second question is whether HFE CFT behavior can be observed before heating occurs in a generic interacting model. To address this, let us specifically ask about perturbative heating rates as discussed in Ref. 11. Specifically, consider quenching a monochromatic drive on an interacting CFT,

$$H(t) = \underbrace{H_{CFT}(U \neq 0) + h_B \sigma_0^z \Theta(t)}_{H_f} + h_B \sigma_0^z \Theta(t) \sin(\omega t).$$

At leading order, the system will absorb energy due to the drive at a rate

$$\frac{dE}{dt} \equiv \frac{d\langle H_f \rangle}{dt} \approx 2h_B^2 \omega \sigma(\omega); \quad \sigma(\omega) = \frac{1}{2}\int_{-\infty}^{\infty} \langle[\sigma_0^z(t), \sigma_0^z(0)]\rangle.$$

Note that this expectation value is intended to be taken in the ground state of the post-quench Hamiltonian $H_f$ which, in the case where heating is slower than CFT dynamics, should be a valid description of heating due to drive near the boundary (where the system looks like it's in the ground state). The question is whether the time scale of this heating is larger than the time scale $t_B \sim h_B^{-\nu_B}$ for the CFT quench physics to take place. For local drive, Abanin et al [11] bound the susceptibility as

$$\sigma(\omega) \lesssim e^{-C\omega/\Lambda},$$

where $C$ is a constant of order 1. An important subtlety is that this involves integrating over a windows of width $\delta\omega$ which is left out of the final expression. From Fermi's golden rule, the relevant width should be the inverse density of states at energy $\omega$ above the ground state, since the argument involves bounding excitation rates from the ground state to states at energy $E_{gs} + \omega$. Therefore, we expect this bound could be tightened, but for now we can substitute this result into the heating rate in order to lower-bound the time scale $\tau_{heat}$ for heating processes to occur.

The most relevant energy scale is the single-particle bandwidth, so let us define $\tau_{\text{heat}} = \Lambda/(dE/dt)$. Then

$$\frac{\tau_{\text{heat}}}{t_B} \gtrsim \frac{\Lambda}{h_B^{2-\nu_B}\omega} e^{C\omega/\Lambda}.$$

For our case, $\nu_B = 2$ and for sufficiently large $\omega/\Lambda$ heating will occur exponentially slower than CFT dynamics. However, for models with sufficiently weak boundary perturbations ($\nu_B > 2$), this ratio will vanish in the scaling limit and the CFT dynamics becomes irrelevant.



\end{document}